\documentclass[%
 aip,
 amsmath,amssymb,
 reprint,%
]{revtex4-1}

\usepackage{graphicx}% Include figure files
\usepackage{dcolumn}% Align table columns on decimal point
\usepackage{bm}% bold math
\usepackage{bm} % bold math
\usepackage[colorlinks=true,
            linkcolor=blue,
            urlcolor=blue,
            citecolor=blue]{hyperref}
\usepackage[utf8]{inputenc}
\usepackage[T1]{fontenc}
\usepackage{mathptmx}
\usepackage{etoolbox}
\usepackage{physics}
\usepackage{lipsum}

\makeatletter
\def\@email#1#2{%
 \endgroup
 \patchcmd{\titleblock@produce}
  {\frontmatter@RRAPformat}
  {\frontmatter@RRAPformat{\produce@RRAP{*#1\href{mailto:#2}{#2}}}\frontmatter@RRAPformat}
  {}{}
}%
\makeatother

\makeatletter
\newcommand{\vast}{\bBigg@{4}}
\newcommand{\Vast}{\bBigg@{5}}
\makeatother

\begin{document}

    \preprint{AIP/123-QED}

    \title{Shrinking shrimp-shaped domains and multistability in the dissipative asymmetric kicked rotor map}
    % Force line breaks with \\ 
\author{Matheus Rolim Sales}
\email{matheusrolim95@gmail.com}
\affiliation{Department of Physics, S\~ao Paulo State University (UNESP), 13506-900, Rio Claro, SP, Brazil}
\author{Michele Mugnaine}
%\affiliation{Department of Physics, Federal University of Paraná, 80060-000, Curitiba, PR, Brazil}
\affiliation{Institute of Physics, University of S\~ao Paulo, 05508-900, S\~ao Paulo, SP, Brazil}
\author{Edson Denis Leonel}
\affiliation{Department of Physics, S\~ao Paulo State University (UNESP), 13506-900, Rio Claro, SP, Brazil}
\author{Iber\^e L. Caldas}
\affiliation{Institute of Physics, University of S\~ao Paulo, 05508-900, S\~ao Paulo, SP, Brazil}
\author{José D. Szezech Jr.}
\affiliation{Graduate Program in Sciences/Physics, State University of Ponta Grossa, 84030-900, Ponta Grossa, PR, Brazil}
\affiliation{Department of Mathematics and Statistics, State University of Ponta Grossa, 84030-900, Ponta Grossa, PR, Brazil}
    \date{\today}% It is always \today, today,
                %  but any date may be explicitly specified

    \begin{abstract}
        
    An interesting feature in dissipative nonlinear systems is the emergence of characteristic domains in parameter space that exhibit periodic temporal evolution, known as shrimp-shaped domains. We investigate the parameter space of the dissipative asymmetric kicked rotor map and show that, in the regime of strong dissipation, the shrimp-shaped domains repeat themselves as the nonlinearity parameter increases while maintaining the same period. We analyze the dependence of the length of each periodic domain with the nonlinearity parameter, revealing that it follows a power law with the same exponent regardless of the dissipation parameter. Additionally, we find that the distance between adjacent shrimp-shaped domains is scaling invariant with respect to the dissipation parameter. Furthermore, we show that for weaker dissipation, a multistable scenario emerges within the periodic domains. We find that as the dissipation gets weaker, the ratio of multistable parameters for each periodic domain increases, and the area of the periodic basin decreases as the nonlinearity parameter increases.

    \end{abstract}

    \maketitle

    \begin{quotation}
        The study of the parameter space of dissipative nonlinear dynamical systems reveals a rich and complex structure between periodic and chaotic behavior. One particularly interesting structure observed in these systems is the so-called shrimp-shaped domains. They are periodic regions embedded in chaos and have been identified in a variety of dynamical systems as well as in experimental realizations. In this paper, we study the dissipative asymmetric kicked rotor map and examine the formation of these shrimp-shaped domains as both the nonlinearity and asymmetry parameters are varied. Our results reveal a scaling relation between the length of the shrinking shrimp-shaped domains and the nonlinearity parameter in the regime of strong dissipation as well as the emergence of multistable scenarios for specific parameter values as the dissipation gets weaker.
    \end{quotation}

    \section{Introduction}
    \label{sec:introduction}
    
    In general, dissipative nonlinear dynamical systems depend on different parameters that influence and determine the nature of their solutions. As these parameters change, the system can undergo transitions known as bifurcations. This topic has been extensively studied in both theoretical \cite{Linsay1981, Parlitz1987,Gilmore1995,Gallas1996,DeOliveira2013, Dacosta2017,Rajagopal2019,Jafari2021, Nkounga2023, Xia2023} and experimental frameworks \cite{Noszticzius1989, Cusumano1995, Valling2007, Linaro2010, Bureau2014, DePaula2017}. We refer the reader to Refs. \cite{ChowHale1982, Crawford1991,Kuznetsov1995} for a detailed discussion on bifurcation theory and applications. A significant discovery was made by Feigenbaum \cite{Feigenbaum1978, Feigenbaum1979} in the late 1970s. He identified a universal constant, now known as the Feigenbaum constant ($\delta \approx 4.669201609$), that characterizes the route to chaos via period-doubling bifurcations. Although this constant was initially derived for one-dimensional mappings, it is found in a wide range of dynamical systems \cite{Hanias2009, Oliveira2011,Chen2012,Korneev2023}, hence its universal character.

    When analyzing bifurcations in two-dimensional parameter spaces, an interesting feature of dissipative systems emerges: the so-called shrimp-shaped domains \cite{Gallas1993, Gallas1994}. These are self-similar, isoperiodic stable structures consisting of periodic windows embedded into quasi-periodic or chaotic regions. While the existence of self-similar periodic structures had been noted in different systems \cite{Gaspard1984,Mackay1986,Rossler1989,Komuro1991}, it was the pioneering work of Gallas \cite{Gallas1993} on the parameter space of the Hénon map that brought significant attention to these shrimp-shaped domains. Since then, these domains have been observed in multiple mathematical models, such as a $\mathrm{CO}_2$ laser model \cite{Bonatto2005}, the Rössler system \cite{Castro2007}, the kicked logistic map \cite{Baptista1997}, a 3D generalization of the Hénon map \cite{Hampton2022}, a two-gene system \cite{Desouza2012}, and a non-ideal Duffing oscillator \cite{Desouza2017}, to cite a few. The shrimp-shaped domains have also been identified in experimental studies involving electronic circuits \cite{Maranhao2008, Stoop2010, Viana2010}. The plethora of dynamical systems that exhibit such a peculiar structure leads to the belief that they are a universal feature of dissipative dynamical systems.

    One dissipative nonlinear system that has been extensively studied is the dissipative version of the standard map (also known as the Chirikov-Taylor map) \cite{Chirikov1979} and referred to as the dissipative standard mapping \cite{Zaslavsky1978}. Despite its simple mathematical form, this system displays all the characteristics typical of nonlinear dynamical systems. Initial studies focused on the transition from quasi-periodicity to chaos \cite{Feigenbaum1982, Rand1982, Ostlund1983, Bohr1984} and from Hamiltonian to dissipative dynamics \cite{Schmidt1985}. The shrimp-shaped domains have been identified in the dissipative standard mapping \cite{Martins2008, Oliveira2011} and the thresholds at which invariant attractors break have been calculated \cite{Calleja2010}. Researchers have also addressed the transport of chaotic particles, including the identification of superdiffusion \cite{Zaslavsky2008} and the identification of a universal empirical function for describing chaotic particle diffusion under weak dissipation \cite{Oliveira2012}. Further research has explored generalizations of this mapping, including fractional \cite{Tarasov2010} and relativistic versions \cite{Ciubotariu2002, Lan2008, Oliveira2011b, Horstmann2017}. Additionally, some analytical analyses have been conducted \cite{Celletti1998, Bustamante2019}.

    In this paper, we consider the asymmetric version of the dissipative standard mapping \cite{Wang2007}. It has been proposed for the study of the transport of particles and the addition of asymmetry leads to a strong ratchet current \cite{Wang2007, Celestino2011, Celestino2014}, which is the directed transport of particles without an external bias force, even for slightly asymmetric potentials \cite{Lopes2012}. Previous studies of this system have revealed a rich structure in the parameter space consisting of the nonlinearity parameter $k$ and the dissipation parameter $\gamma$ \cite{Celestino2011,Celestino2014}. We, on the other hand, seek to describe its two-dimensional parameter space consisting of the nonlinearity parameter $k$ and the asymmetry parameter $a$. We identify a distinctive cascade of shrimp-shaped domains as $k$ grows larger and obtain a scaling relation between their lengths and the parameter $k$. We also find that the distance between adjacent shrimp-shaped domains is scaling invariant with respect to the dissipation. Additionally, we identify specific parameter combinations that result in a multistable scenario within the periodic domains as the dissipation decreases.
    
    %Additionally, we identify several combinations of parameters that lead to a multistable scenario within the periodic domains as the dissipation gets weaker.

    %This paper is organized as follows. In Section \ref{sec:model}, we describe the system under study and present some of its properties. In Section \ref{sec:tinyshrimps}, we analyze the cascade of shrimp-shaped domains in parameter space for large values of the nonlinearity parameter. We analyze the length of the shrimp-shaped structures as a function of the nonlinearity parameter and show that the rate at which their size diminishes does not depend on the dissipation parameter. Additionally, we show that the distance between adjacent shrimp-shaped structures is scaling invariant with respect to the dissipation parameter. Furthermore, we show that for weaker dissipation, the existence of several attractors becomes relevant, and in Section \ref{sec:multi} we identify and analyze the multistability within the periodic structures. We show that the multistability scenario is more prominent for weaker dissipation. We also demonstrate that as the nonlinearity parameter increases, the size of the periodic basins decreases. Section \ref{sec:conclusions} contains our final remarks.

    This paper is organized as follows. In Section \ref{sec:model}, we describe the system under study and present some of its properties. In Section \ref{sec:tinyshrimps}, we analyze the cascade of shrimp-shaped domains in parameter space for large values of the nonlinearity parameter, and in Section \ref{sec:multi}, we identify and analyze the multistability within the periodic structures. We show that the multistability scenario is more prominent for weaker dissipation. We also demonstrate that as the nonlinearity parameter increases, the size of the periodic basins decreases. Section \ref{sec:conclusions} contains our final remarks.
    
    \section{The model}
    \label{sec:model}
    
    In this paper, we study a periodically kicked rotor subjected to an asymmetric harmonic potential. The canonical variables are the momentum $p_n$ and the angular position $x_n$ of the rotor just after the $n$th kick. The dynamics is given by the dissipative asymmetric kicked rotor map (DAKRM) \cite{Wang2007, Celestino2011, Lopes2012}:
    \begin{equation}
        \label{eq:mapping}
        \begin{aligned}
            p_{n + 1} &= (1 - \gamma)p_n + k\qty[\sin(x_n) + a\sin\qty(2x_n + \frac{\pi}{2})],\\
            x_{n + 1} &= x_n + p_{n + 1}\mod2\pi,
        \end{aligned}
    \end{equation}
    where $k > 0$ corresponds to the kick strength, $\gamma \in [0, 1]$ is the dissipation parameter, and for $a \neq 0$, the spatial symmetry is broken. For $a = 0$, the dissipative standard mapping \cite{Feigenbaum1982, Rand1982, Ostlund1983, Bohr1984} is recovered, and the limiting cases $\gamma = 0$ and $\gamma = 1$ correspond to Hamiltonian and overdamping dynamics, respectively.

    \begin{figure}[tb!]
        \centering
        \includegraphics[width=\linewidth]{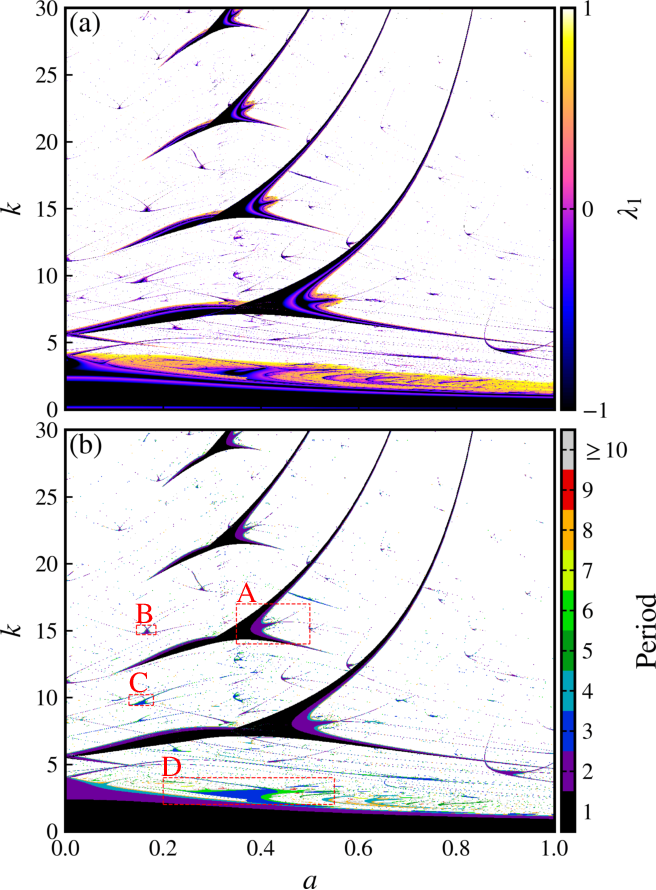}
        \caption{(a) The largest Lyapunov exponent and (b) the isoperiodic diagram in parameter space $k \times a$ with $\gamma = 0.80$. In (a), blue to black (pink to white) color corresponds to periodic (chaotic) dynamics, whereas purple color (vanishing $\lambda_1$) marks the bifurcation points. In (b), the colored regions correspond to different periods in the periodic domains and the white region corresponds to chaotic dynamics. The red dashed rectangles labeled from A to D correspond to the magnifications shown in Fig.~\ref{fig:paramspacezooms}. The initial condition is chosen as $(x_0, p_0) = (1.78, 0.0)$ for all values of $(a, k)$.}
        \label{fig:paramspace}
    \end{figure}

    \begin{figure*}[tb]
        \centering
        \includegraphics[width=\linewidth]{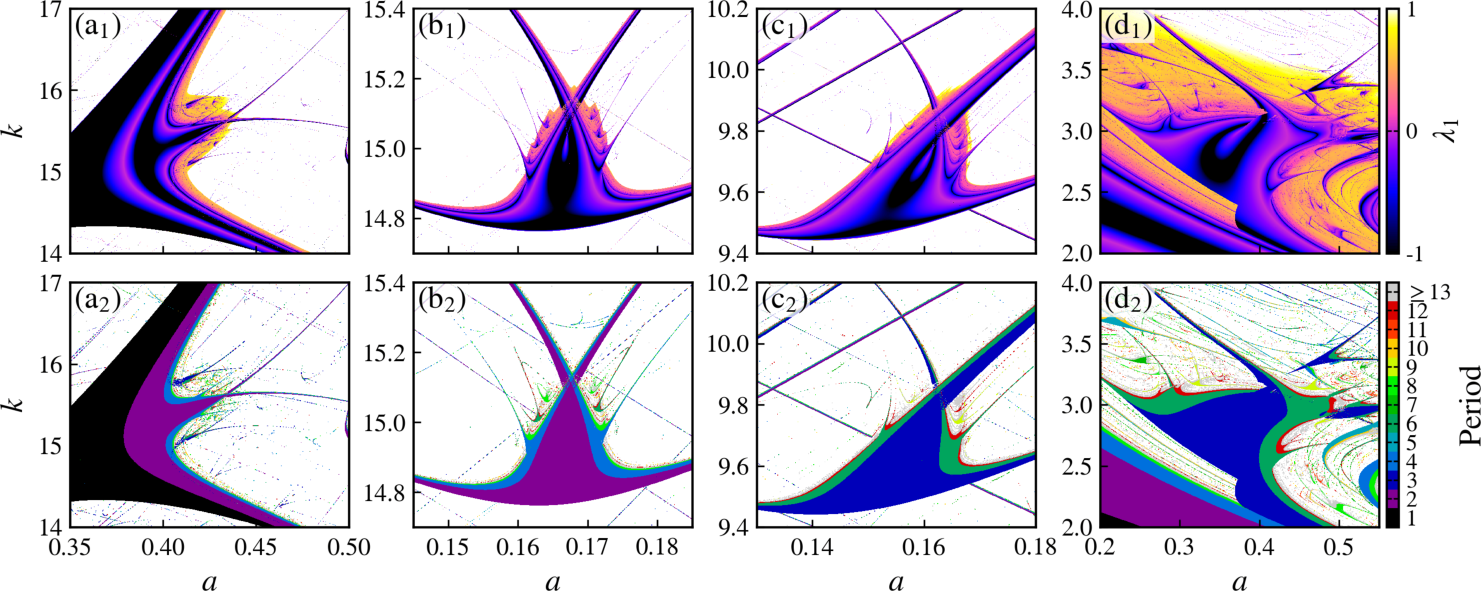}
        \caption{(Top row) The largest Lyapunov exponent and (bottom row) the isoperiodic diagram for the regions bounded by the red dashed rectangles labeled from A to D in Fig.~\ref{fig:paramspace}, respectively. Note that the color code for the period has slightly changed from the one used in Fig.~\ref{fig:paramspace}.}
        \label{fig:paramspacezooms}
    \end{figure*}

    We are interested in the regime of strong dissipation and consider, at first, $\gamma = 0.80$. We analyze the nature of the solutions and the different dynamical behaviors that emerge for different parameter values. In order to do so, we calculate the largest Lyapunov exponent \cite{Shimada1979,Benettin1980,Wolf1985} in the parameter space $k \times a$ for fixed $\gamma$. Given a mapping $\vb{M}: \mathbb{R}^d\rightarrow\mathbb{R}^d$, defined as $\vb{x}_{n + 1} = \vb{M}(\vb{x}_n) = \vb{M}^n(\vb{x}_0)$, let $\vb{DM}^n$ be the $n$ iterate of the Jacobian matrix. The Lyapunov exponents are defined as \cite{Eckmann1985}
    \begin{equation}
        \label{eq:lyapunov}
        \lambda_i = \lim_{n\rightarrow\infty}\frac{1}{n}\ln\qty(\norm*{\vb{DM}^nu_i}),
    \end{equation}
    where $i = 1, 2, \ldots, d$ and $u_i$ is the eigenvector corresponding to the $i$th eigenvalue of $\vb{DM}^n$. In our case, the system is two-dimensional, $d = 2$, and we have two exponents that satisfy $\lambda_1 \geq \lambda_2$. We characterize the dynamics using the largest Lyapunov exponent, $\lambda_1$. The dynamics is periodic when $\lambda_1 < 0$. Chaotic dynamics, on the other hand, is characterized by $\lambda_1 > 0$. Thus, we divide the parameter space $(a, k) \in [0, 1] \times [0, 30]$ into a grid of $1000 \times 1000$ boxes and calculate the largest Lyapunov exponent, $\lambda_1$ [Fig.~\ref{fig:paramspace}(a)]. We also count the period for each pair of points in the grid [Fig.~\ref{fig:paramspace}(b)]. For all our simulations, we consider a fixed initial condition $(x_0, p_0) = (1.78, 0.0)$, a transient time of $5 \times 10^4$ and calculate $\lambda_1$ using the next $5 \times 10^4$ iterations. 
    
    The parameter space in Fig.~\ref{fig:paramspace} shows chaotic (orange to white color) and regular (blue to black color) regions non-trivially intertwined. By changing one parameter while keeping the other fixed, or even by changing both simultaneously, we observe that a chaotic dynamics region shifts to periodic dynamics which then shifts back to chaotic dynamics, thus creating periodic windows. These periodic windows form the so-called shrimp-shaped domains, which are periodic regions in parameter space surrounded by chaotic regions \cite{Mackay1986, Gallas1993,Gallas1994,Baptista1997}. There are also values of $(a, k)$ where $\lambda_1$ approaches zero (purple color). These are bifurcation points and to analyze these bifurcations in further detail, we perform the same simulations for the regions bounded by the red dashed rectangles shown in Fig.~\ref{fig:paramspace}(b). Figure~\ref{fig:paramspacezooms} shows (top row) $\lambda_1$ and (bottom row) the period for the aforementioned regions and we notice that all of these bifurcations are period-doubling bifurcations. Each shrimp-shaped domain consists of a main region followed by an infinite sequence of period-doubling bifurcations that lead to chaotic dynamics. These bifurcations follow the rule $P_0 \times 2^m$, where $P_0$ is the period of the main region. In the case of the larger periodic domains, $P_0 = 1$, however, for smaller periodic domains, there are cases where $P_0 = 2$ [Fig.~\ref{fig:paramspacezooms}(b)] and also $P_0 = 3$ [Figs.~\ref{fig:paramspacezooms}(c) and~\ref{fig:paramspacezooms}(d)]. 

    Furthermore, another interesting feature arises in the parameter space $k \times a$. In Fig.~\ref{fig:paramspace} we notice that the larger shrimp-shaped domains repeat themselves for a seemingly periodic interval in $k$ (at least up to $k = 30$) while their size diminishes. Their position in $a$ is shifted to smaller values, however. For the standard mapping, \textit{i.e.}, for $a = 0$ and $\gamma = 0$, it is known that as the nonlinearity parameter $k$ changes, new elliptical fixed points surrounded by stability islands, which are called ``islets'' due to their small size, appear in approximate intervals of $2\pi$ \cite{Chirikov1979, Manos2014}. Recently, Nieto \textit{et al.} \cite{Nieto2024}, after performing a systematic search for the islets, demonstrated that the length, the area, and the volume of the islets decay following a power law with exponents $-1$, $-2$, and $-3$, respectively, as a function of $k$. Therefore, in Section~\ref{sec:tinyshrimps} we investigate and perform a scaling analysis on the cascade of shrimp-shaped domains shown in Fig.~\ref{fig:paramspace} for larger values of $k$.

    \section{Shrimp-shaped domains scaling analysis}
    \label{sec:tinyshrimps}

\begin{figure}[tb]
        \centering
        \includegraphics[width=\linewidth]{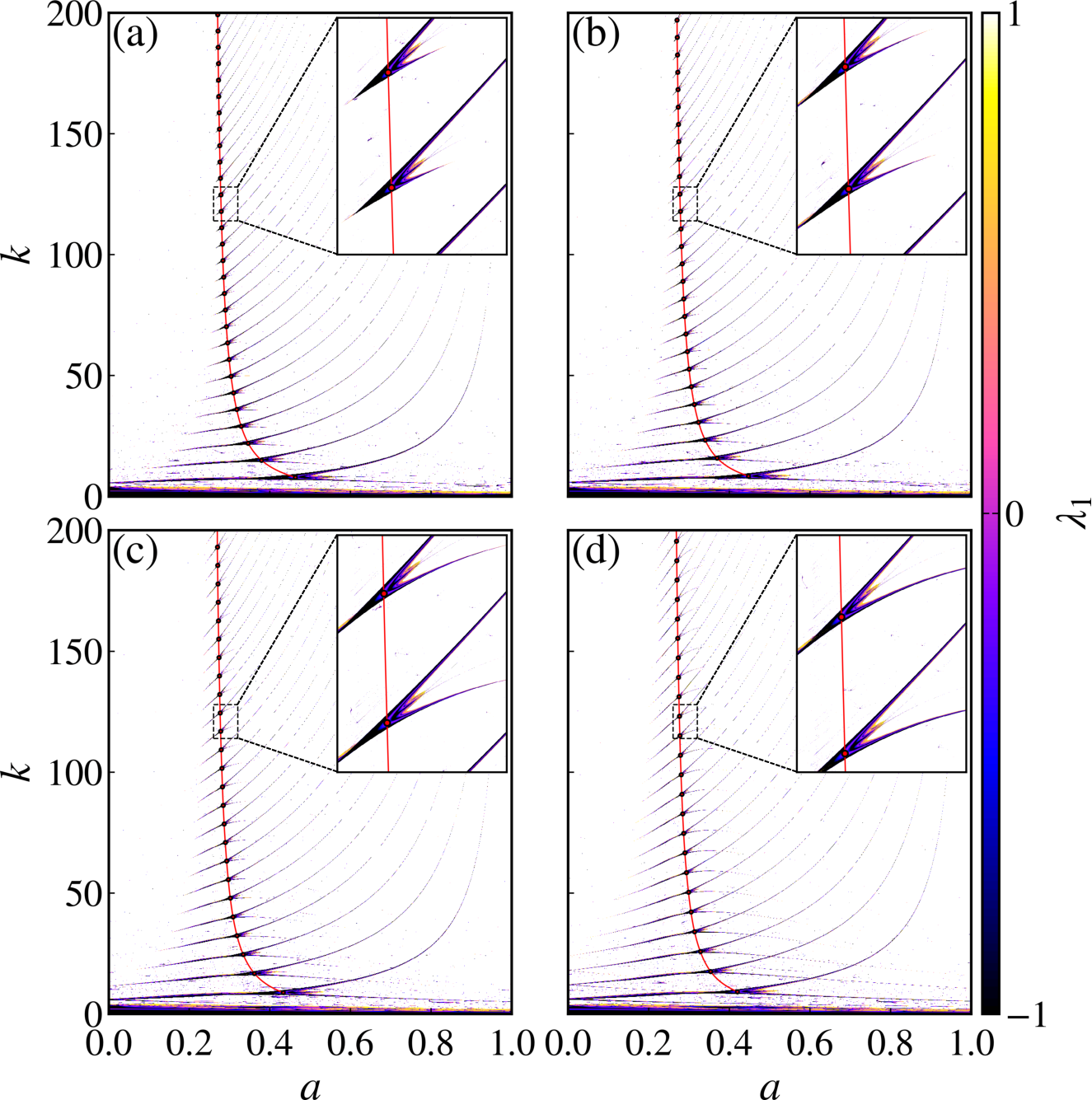}
        \caption{The largest Lyapunov exponent in parameter space $k \times a$ for (a) $\gamma = 0.80$, (b) $\gamma = 0.85$, (c) $\gamma = 0.90$, and (d) $\gamma = 0.95$. The initial condition is chosen as $(x_0, p_0) = (1.78, 0.0)$ for all $(a, k)$. The red curve is a 17th-order polynomial fitting of the red dots at the ``center'' of each shrimp. The set of values of $a$ and $k$ for each $\gamma$ and the corresponding fitting coefficients can be found in the Supplementary Material. The insets are magnifications of the regions bounded by the white dashed rectangles.}
        \label{fig:paramspacetinyshrimp}
    \end{figure}

    In this section, we study the parameter space $k \times a$ for larger values of $k$ considering different values for the dissipation parameter. We calculate $\lambda_1$ in the parameter space $(a, k) \in [0, 1] \times [0, 200]$ for $\gamma = 0.80, 0.85, 0.90,\text{ and } 0.95$ (Fig.~\ref{fig:paramspacetinyshrimp}) following the procedure discussed in Section~\ref{sec:model}. The cascade of shrimp-shaped domains shown in Fig.~\ref{fig:paramspace} indeed persists for large values of $k$ and the larger the value of $k$, the smaller the periodic domain. The insets in Fig.~\ref{fig:paramspacetinyshrimp} show magnifications of some smaller periodic domains. To analyze the dependence of the size of the periodic domains as $k$ varies, we cannot simply fix one value of $a$ and vary $k$ as the domains shift to smaller values of $a$ and this would lead to a biased measurement of the length of each domain. Therefore, for each shrimp-shaped domain in Fig.~\ref{fig:paramspacetinyshrimp}, we define a point $(a, k)$ as its ``center'' (red dots in Fig.~\ref{fig:paramspacetinyshrimp}). We perform a 17th-order polynomial fitting with $a$ as a function of $k$ [$a = a(k)$] (red curve in Fig.~\ref{fig:paramspacetinyshrimp}) and analyze the length of the periodic domains along this function with $a$ and $k$ changing simultaneously according to $a = a(k)$. The values of $a$ and $k$ as well as the obtained coefficients are available in the Supplementary Material.

    \begin{figure}[tb]
        \centering
        \includegraphics[width=\linewidth]{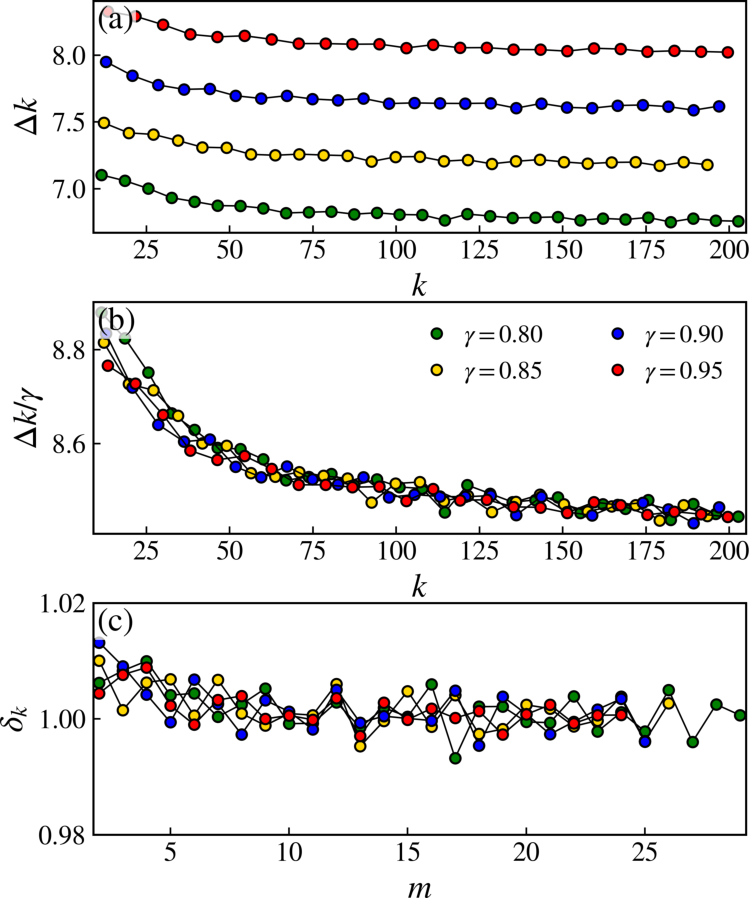}
        \caption{(a) The distance between two successive values of $k$ corresponding to the centers of the shrimp-shaped domains, \textit{i.e.}, $\Delta k = k_{m + 1} - k_m$ as a function of $k$ for the values of $\gamma$ indicated in (b). (b) The overlap of the curves after the transformation $\Delta k \rightarrow \Delta k/\gamma$, indicating the distance between adjacent shrimp-shaped domains as a function of $k$ is scale-invariant with respect to $\gamma$. (c) The Feigenbaum-like constant [Eq. \eqref{eq:feigen}] for $k$.}
        \label{fig:ratios}
    \end{figure}

    However, let us first analyze the obtained positions $(a, k)$. Figure~\ref{fig:ratios}(a) shows the distance in $k$ between adjacent centers as a function of $k$. The larger the dissipation parameter $\gamma$, the larger the interval between two shrimp-shaped domains. Also, for all values of $\gamma$, the distance $L$ is large for small $k$ and saturates to an approximately constant value as $k$ increases. Additionally, we note that even though $\Delta k(\gamma)$ has different values for different $\gamma$, the behavior of the curves is the same. Indeed, by employing the transformation $\Delta k\rightarrow \Delta k/\gamma$, the curves overlap into a single, and hence, universal curve [Fig.~\ref{fig:ratios}(b)]. This indicates scaling invariance \cite{Leonel2004, Leonel2007} of the distance between two shrimp-shaped domains with respect to $\gamma$. When a quantity in a dynamical system exhibits scaling invariance, its expected behavior remains consistent and robust across different scales. This means the system can be rescaled such that, after appropriate parametrization, the quantity remains scale-independent and displays universal features\cite{LeonelBook}. We also compute a Feigenbaum-like constant, defined as:
    \begin{equation}
        \label{eq:feigen}
        \delta_k = \lim_{m\rightarrow\infty}\frac{k_{m - 1} - k_{m - 2}}{k_{m} - k_{m - 1}},
    \end{equation}
    for the values of $k$ corresponding to the centers of the shrimp-shaped domains [Fig.~\ref{fig:ratios}(c)]. This constant is used to describe the universal route to chaotic motion via period-doubling bifurcations in one-dimensional mappings \cite{Feigenbaum1978,Feigenbaum1979}. In our case, we are not analyzing bifurcations but rather the separation of adjacent shrimp-shaped domains in $k$ and obtain $\delta_k \rightarrow 1$ as $m \rightarrow \infty$, which is another indication that the distance of adjacent shrimp-shaped domains tends to a constant value as $k$ increases.
    
    \begin{figure*}
        \centering
        \includegraphics[width=\linewidth]{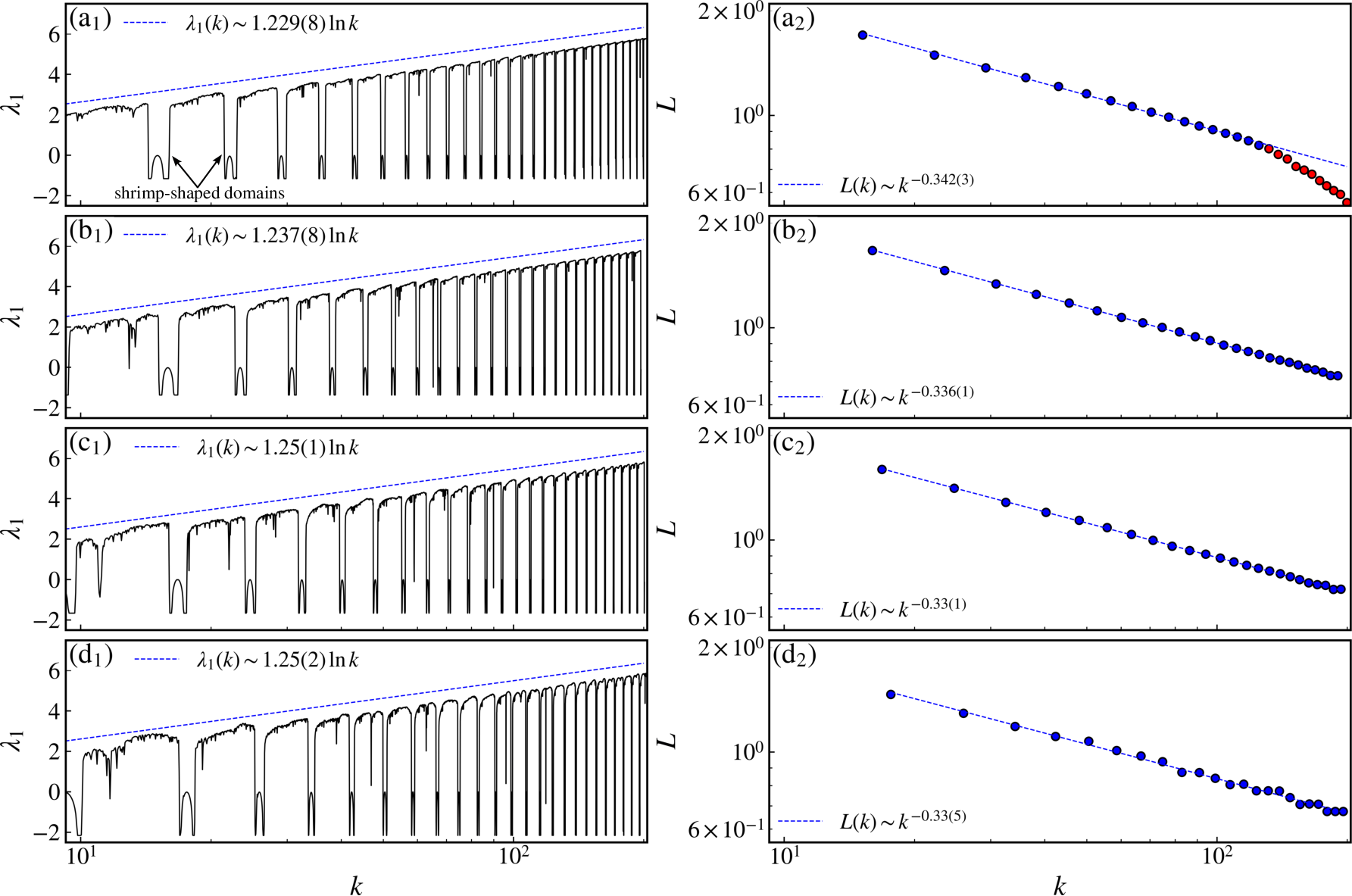}
        \caption{(Left column) The largest Lyapunov exponent and (right column) the length of the periodic windows as a function of $k$ for (a) $\gamma = 0.80$, (b) $\gamma = 0.85$, (c) $\gamma = 0.90$, and (d) $\gamma = 0.95$. As $k$ changes, the parameter $a$ also changes according to the function $a = a(k)$ (red curves in Fig.~\ref{fig:paramspacetinyshrimp}).}
        \label{fig:lyapak}
    \end{figure*}

    \begin{figure}
        \centering
        \includegraphics[width=\linewidth]{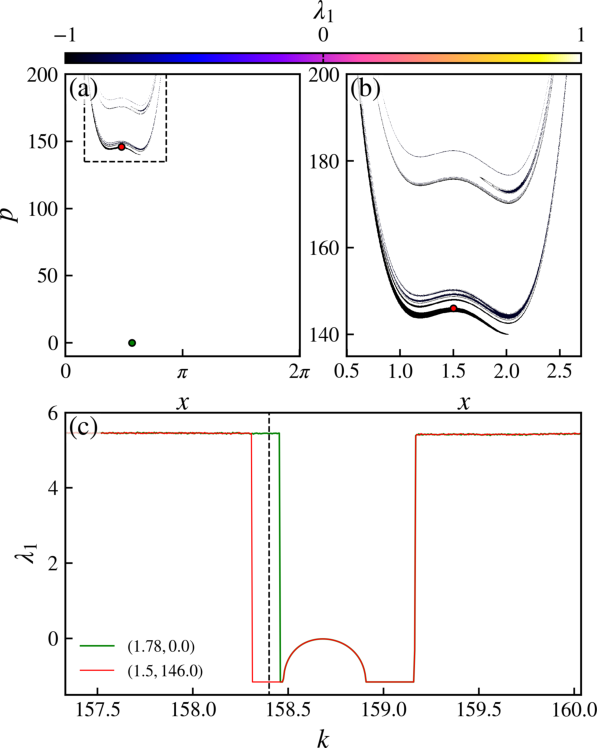}
        \caption{(a) The largest Lyapunov exponent for a grid of $1000 \times 1000$ initial conditions uniformly distributed in the phase space region $(x, p) \in [0, 2\pi) \times [0, 200]$ with $k = 158.4$ and $a \approx 0.27372$ [calculated according to $a = a(k)$] and (b) is a magnification of the region indicated by the black dashed lines. (c) The largest Lyapunov exponent as a function of $k$ and $a$ [$a = a(k)$] for two distinct initial conditions indicated in (a) by the colored dots. The black vertical dashed line indicates the values of $k$ used in (a).}
        \label{fig:multi}
    \end{figure}    

    In order to determine the size of the shrimp-shaped domains, we calculate $\lambda_1$ as a function of $k$ and $a$ simultaneously, where $a = a(k)$ for the same values of $\gamma$ used in Figs.~\ref{fig:paramspacetinyshrimp}~and~\ref{fig:ratios} (left column of Fig.~\ref{fig:lyapak}). We notice several windows where $\lambda_1 < 0$. These periodic windows correspond to the shrimp-shaped domains in the parameter space in Fig.~\ref{fig:paramspacetinyshrimp}. For each of these windows, we determine their length $L$ as a function of $k$ (right column of Fig.~\ref{fig:lyapak}). The power law dependence is evident (blue dots) and the exponents, within numerical errors, are the same which indicates that the rate at which the periodic domains shrink with $k$ does not depend on $\gamma$. Additionally, the value of $\lambda_1$ between two adjacent periodic windows scales with $k$ as $\sim C \times \ln{k}$, with $C$ being approximately the same in all cases (within numerical errors). Therefore, in the regime of strong dissipation, both the length of the periodic domains and the magnitude of $\lambda_1$ within the chaotic regions do not depend on the dissipation parameter $\gamma$ but rather exhibit approximately equal scaling relations for all $\gamma$.

    Regarding the deviation from the power law decay observed in Fig.~\ref{fig:lyapak}(a$_2$) (red dots), it arises due to the presence of several attractors for specific parameter values. Figures~\ref{fig:multi}(a)~and~\ref{fig:multi}(b) display $\lambda_1$ for a grid of initial conditions in phase space with fixed values of $k$ and $a(k)$. Most of the phase space corresponds to chaotic dynamics, with a few periodic regions coexisting within the large chaotic region. Note that the initial condition we have been using, $(x_0, p_0) = (1.78, 0.0)$, indicated by the green dot, lies in the chaotic region for these specific parameter values. We fix this initial condition and vary $k$ in the interval $k \in [157.5, 160]$, then calculate $\lambda_1$ [green curve in Fig.~\ref{fig:multi}(c)]. We observe a slightly different behavior by comparing this curve with those in Fig.~\ref{fig:lyapak}. For instance, consider the first periodic window in Fig.~\ref{fig:lyapak}(a$_1$). As $k$ changes, $\lambda_1$ initially remains positive but then suddenly drops to $\lambda_1 < 0$, indicating the beginning of the periodic domain. The value of $\lambda_1 < 0$ remains constant over an interval in $k$, after which it smoothly rises towards zero and then drops back to its previous negative value. With a slight further change in $k$, $\lambda_1$ abruptly increases to a positive value, marking the end of the periodic domain. In the case of the green curve in Fig.~\ref{fig:multi}(c), the initial region where $\lambda_1$ remains negative after the sudden transition from positive to negative $\lambda_1$ seems shorter than other cases in Fig.~\ref{fig:lyapak}. Indeed, by choosing an initial condition that belongs to the periodic region for $k = 158.4$ [red dot in Fig.~\ref{fig:multi}(a)] and calculating $\lambda_1$ for the same interval in $k$ [red curve in Fig.~\ref{fig:multi}(c)], we recover the observed behavior in Fig.~\ref{fig:lyapak}. Thus, the values of $L$ corresponding to the red dots in Fig.~\ref{fig:lyapak}(a$_2$) are actually the length of ``incomplete'' periodic windows, illustrated by the green curve in Fig.~\ref{fig:multi}, when, in fact, the ``true'' periodic window is illustrated by the red curve. This results in an incorrect measurement of the shrimp-shaped domain length and causes the deviation from the power law decay. Fixing this measurement is no easy task as the basin of attraction of the periodic region strongly depends on the parameters $k$ and $a$.
    
    Therefore, due to the existence of many attractors for specific parameter values, different initial conditions should yield similar, but not the same, parameter spaces, and this feature is further analyzed in Section~\ref{sec:multi}.
        
    \section{Multistability}
    \label{sec:multi}
    
    In many systems, there is more than only one asymptotic state or attractor, indicating a multistability scenario where the final state of a trajectory depends strongly on its initial condition \cite{Feudel1997}. Multistability is characterized by the coexistence of different attractors for a certain set of parameters, just as shown in Figure~\ref{fig:multi}(a) where for fixed values of $a$ and $k$, we identify a chaotic attractor with a tenuous red basin and also a periodic attractor, with a blue basin of attraction.

\begin{figure}
    	\centering
    	\includegraphics[width=\linewidth]{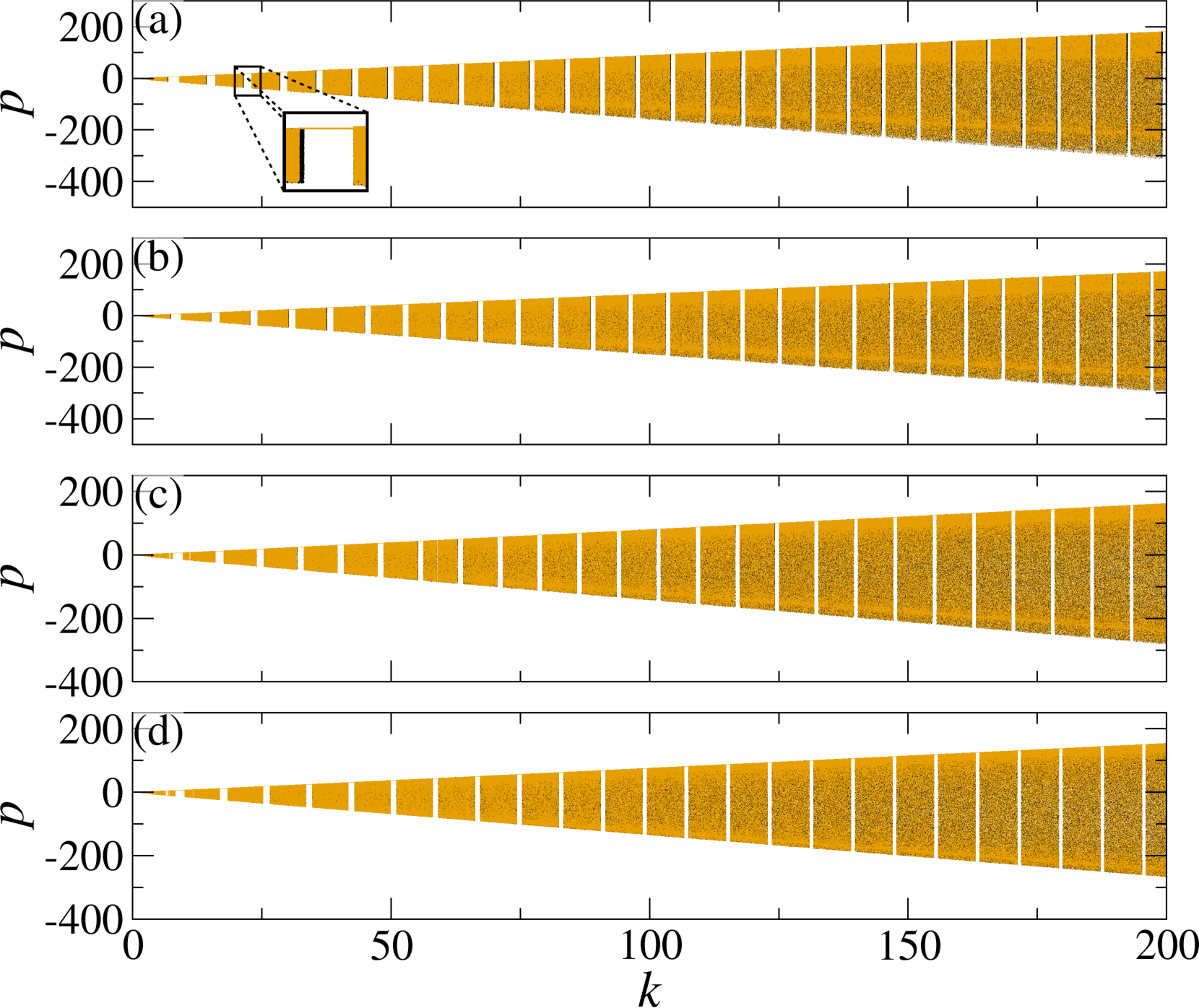}
    	\caption{Existence of hysteresis on the bifurcation diagrams computed on the curve $k[a=a(k)]$ (red curves in Fig~\ref{fig:paramspacetinyshrimp}) and the four values of $\gamma$: (a) $\gamma=0.80$, (b) $\gamma=0.85$, (c) $\gamma=0.90$ and (d) $\gamma=0.95$. The black (orange) points indicate the diagram calculated for increasing (decreasing) values of $k$. The hysteresis and, consequently, the multistability is highlighted in the amplification of panel (a).}
    	\label{fig:Diagram}
    \end{figure}

    One method to identify the parameters related to multistability is by the hysteresis of the bifurcation diagrams, \textit{i.e.}, we compute the diagram for two directions of variation of the parameter. Here, we choose to compute the diagram in the curve $a(k)$ exhibited as the red points in Figure~\ref{fig:paramspacetinyshrimp}, following the attractor. The two chosen directions are the increase and decrease of $k$. For each direction, we plot the diagram with different colors and then superimposed them in order to identify regions of multistability. The results on the bifurcation diagrams for $\gamma=0.8, 0.85, 0.9$, and 0.95 are shown in Fig.~\ref{fig:Diagram}.

In Fig.~\ref{fig:Diagram}, the black (orange) points indicate the diagram computed for increasing (decreasing) values of $k$. The diagrams are superimposed and for all chaotic regions we have black and orange points coexisting. For all panels, we observe a domain where periodic windows separate the chaotic regions. The sequence of periodic windows is in accordance with Fig.~\ref{fig:lyapak}. Observing and comparing all the panels, the value of $\gamma$ does not significantly affect the bifurcations diagrams, since they are visually similar for all the $\gamma$ analyzed.

\begin{figure}
    	\centering
    	\includegraphics[width=\linewidth]{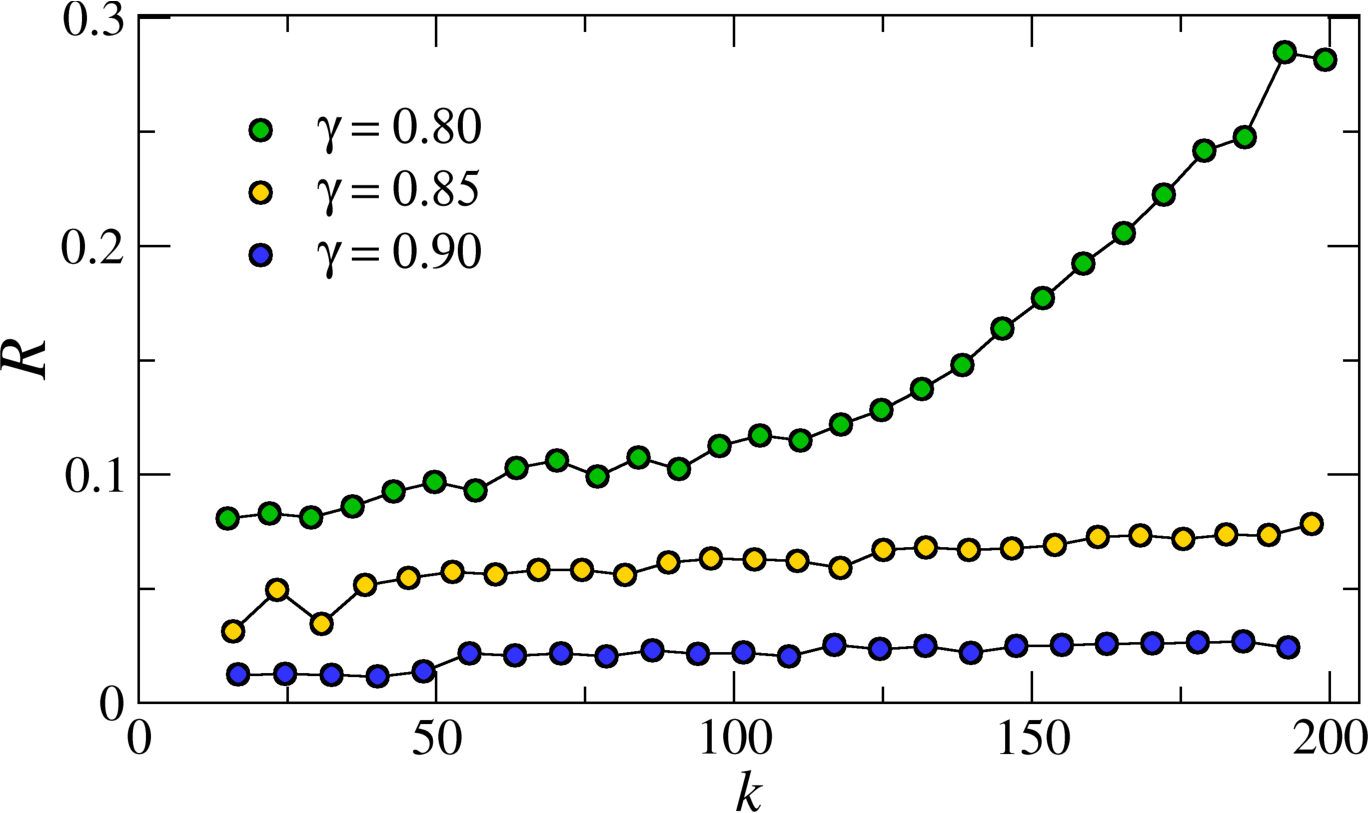}
    	\caption{The ratio of multistable parameters for each periodic window in the bifurcation diagram of Fig. \ref{fig:Diagram} (a)-(c). The $k$-coordinate refers to the value of $k$ of the center of each shrimp-shaped domain related to the periodic window. We have not found multistability for $\gamma=0.95$.}
    	\label{fig:RatioMS}
    \end{figure}

In order to analyze the multistability inside the periodic windows, we highlight the region with the amplification in Fig.~\ref{fig:Diagram}(a). As we observe, the multistability scenario is formed by the coexistence of chaotic attractor, represented by the black points in this interval, with the fixed point in orange. This scenario repeats for all the periodic windows in the diagram, for all studied values of $\gamma$. From our computations of the bifurcation diagrams shown in Fig.~\ref{fig:Diagram}, there is a subtle decrease in the size of the black chaotic regions in the multistable intervals as $\gamma$ increases.  In order to verify this implication, we compute the ratio $R$ of parameters with multistability in each periodic window. The result is shown in Fig. \ref{fig:RatioMS}.

The ratio $R$ presented in Fig.~\ref{fig:RatioMS} is the ratio between the range of parameters where there is multistability by the range of the entire periodic window. The value of $k$ of each point is the value of the center of the corresponding shrimp-shaped structure. For $\gamma=0.8$, we observe that as $k$ increases, the ratio of parameters with multistability also increases, up to a scenario where almost $30\%$ of the periodic window presents multistability. For greater values of $\gamma$, $\gamma=0.85$, and $\gamma=0.90$, the ratio of multistable parameters is smaller than $10\%$ for all periodic windows and, as $k$ increases, the ratio $R$ does not increase significantly, compared to the increase for $\gamma=0.8$. Therefore, the increase in the dissipation amplitude decreases the multistability regions in the bifurcation diagrams. This result is in accordance with the studies presented in \cite{Feudel1996,Feudel1998}, where the presence of small dissipation can lead to an arbitrarily large number of attractors coexisting in the system.

Another method to describe the coexistence of attractors and, consequently, the coexistence of different basins of attractors is by the basin stability. The basin stability can be defined as the quantification of the volume of the basin \cite{Menck2013}. According to the authors, the basin's volume is related to the probability of returning to its state after any random perturbation. Thus, the greater the basin, the more stable it is. 

\begin{figure*}[tb]
   	\centering
   	\includegraphics[width=0.95\linewidth]{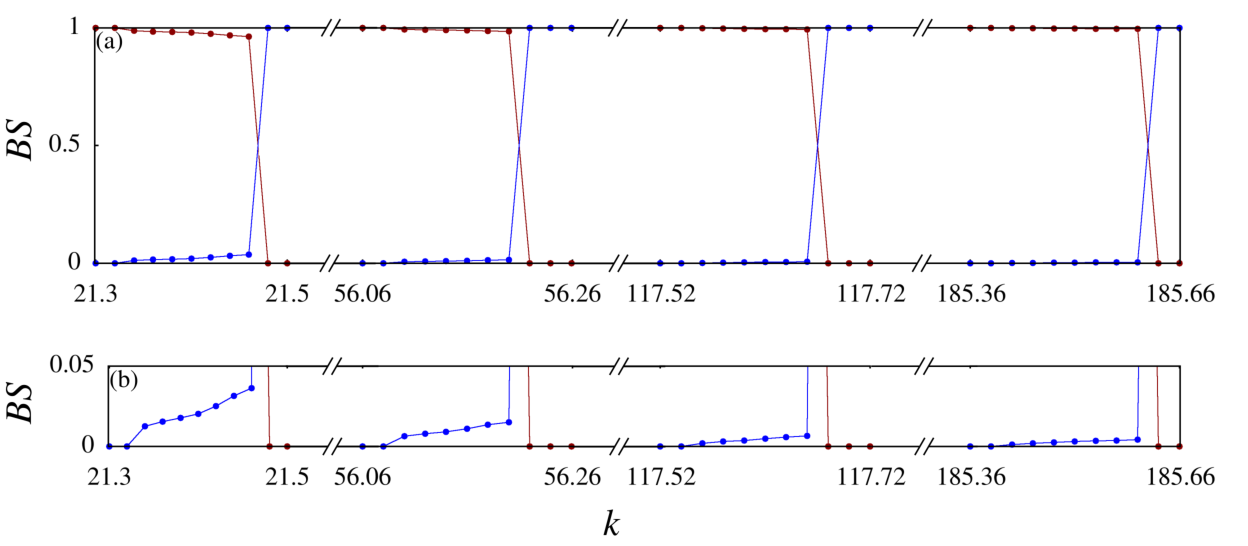}
   	\caption{Basin stability ($BS$) as a function of $k$ computed for $\gamma=0.8$ over the red curve of Fig.~\ref{fig:paramspacetinyshrimp}(a) for four intervals with multistability. We present all the intervals in the same graph by breaking the $k-$ axis.  The blue (red) points indicate the value of $BS$ for the periodic (chaotic) basin. In (a), we can observe the transitions from a single chaotic basin to a periodic basin. In (b), we amplify the graph for small values of $BS$ in order to enhance the increase in the $BS$ of the periodic basin.}
   	\label{fig:basinStability}
\end{figure*}

From the diagrams shown in Fig.~\ref{fig:Diagram}, we observe two possible attractors: periodic or chaotic. Thus, the basin stability for such basins will not be null or equal to unity in intervals where the system presents multistability. In order to analyze basin stability in multistability scenarios for the DAKRM, we choose four periodic windows of Fig.~\ref{fig:Diagram}(a) and compute the respective fraction of the area occupied by each basin, \textit{i.e.}, the stability of the basin (Fig.~\ref{fig:basinStability}).

The basin stability for each basin is computed by the analysis of $10^6$ points uniformly distributed in the phase space $(x, p)\in[0,2\pi]\times[0,200]$. We iterate each point for $10^3$ iterations and compute the fixed points (if they exist) to each parameter in the intervals. If the solution converges to a periodic attractor (fixed point), the point belongs to the periodic basin, otherwise, the point belongs to the chaotic basin.

We compute the basin stability for four different intervals of multistability: (1) $k\in[21.3,21.5]$, (2) $k\in[56.06,56.26]$, (3) $k\in[117.52,117.72]$ and (4) $k\in[185.36,186.66]$. In order to present all intervals in the same graph, we break the $k$-axis into different intervals, and the breaks are indicated by the $//$ symbols. From the basin stability behavior shown in Fig.~\ref{fig:basinStability}(a), we observe that multistable cases (where $BS\ne 0$ and $BS\ne1$) are transitions from a scenario with just chaotic attractor (red points in $BS=1$) to a scenario with just a periodic basin (blue points in $BS=1$). For all intervals, we observe a smooth and subtle decrease in the $BS$ of the chaotic basin followed by an abrupt decrease to $BS=0$. Since we only have two possible basins, $BS_{chaotic}+BS_{periodic}=1$. Thus, the basin stability for the periodic basin increases abruptly or smoothly as the chaotic basin changes.

\begin{figure}[!h]
   	\centering
   	\includegraphics[width=\linewidth]{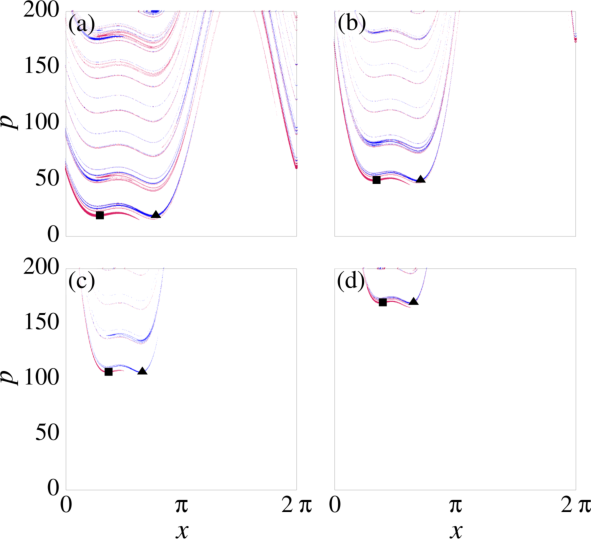}
   	\caption{Attraction basin for multistability cases. For all studied cases, we have the coexistence of a chaotic attractor (not shown in the picture) and two fixed point attractors. The chaotic basin is shown in red, while the basin related to the periodic attractors is shown in blue and yellow. The values of $k$ are (a) $k=21.46 $, (b) $k=56.2$, (c) $k=117.66$, (d) $k=185.6$, and, for all panels, $\gamma=0.8$.}
   	\label{fig:Basins}
\end{figure}

Observing the magnification in Fig.~\ref{fig:basinStability}(b), as the value of $k$ increases, the basin stability for the periodic basin increases up to lower values until it becomes the only basin in the system. Thus, the area of the periodic basin in a multistable scenario decreases as $k$ increases. In order to illustrate this result, we choose, for all intervals in Fig.~\ref{fig:basinStability}, the last parameter $k$ with multistability and compute the respective basins of attraction (Fig.~\ref{fig:Basins}).

From the basin of attractions shown in Fig.~\ref{fig:Basins}, we observe that as $k$ increases, the area of the periodic basin decreases, for the chosen parameters. Besides, the $p$ value of the attractors increases as $k$ assumes greater values. These two consequences affect the computation of the Lyapunov exponent. Since the periodic basin shrinks as $k$ increases and the value of $p$ changes for every periodic window, it is less likely to choose one periodic initial condition to compute $\lambda$.  With this, we indeed verify that the deviation represented by the red points in Fig.~\ref{fig:lyapak}(a$_2$) is a consequence of multistability. This deviation is not seen for greater values of $\gamma$ because as $\gamma$ increases, there are fewer multistable regions and they are smaller, implying less influence in the Lyapunov exponent computation.
\section{Conclusions}
    \label{sec:conclusions}

We have performed a scaling analysis in the parameter space of the dissipative asymmetric kicked rotor map and explored multistable scenarios for different parameter values. Our initial analysis focused on the parameter space $k \times a$ in the regime of strong dissipation. We have found a cascade of shrimp-shaped domains in which these periodic structures repeat themselves for increasing values of $k$ and the larger the values of $k$, the smaller the periodic domain. We have demonstrated that the length of the shrimp-shaped domains scales with $k$ as a power law and the decay exponent does not depend on the dissipation parameter $\gamma$. Additionally, we have found that as the dissipation becomes weaker and the nonlinearity parameter grows larger, the existence of several attractors becomes relevant. This results in an incorrect measurement of the length of the shrimp-shaped domains due to the reduction in the size of the periodic basin, leading to a deviation in the scaling of the lengths of these domains.

When analyzing the multistability scenarios, we have observed that inside the periodic windows in the bifurcation diagram, there is a narrow region with the coexistence of chaotic and periodic attractors. This region decreases and eventually disappears for greater dissipation parameters $\gamma$. By computing the ratio of the multistable region for each periodic window, we have identified an increase of the ratio as $k$ increases when considering weaker dissipation, whereas, for greater dissipation, the ratio is almost constant. We have also calculated the basin stability and the basins of attraction. We have demonstrated that as $k$ increases, the area occupied by the periodic basin decreases and is restricted to larger values of the coordinate $p$ around the fixed points. These two results explain the incorrect measurement of shrimp-shaped properties, since the periodic behavior becomes more rare and more restricted to some initial conditions in the system.

%When analyzing the multistability scenarios, we observe that inside the periodic windows in the bifurcation diagram, there is a narrow region with the coexistence of chaotic and periodic attractors. This region decreases and eventually disappears for greater dissipation parameters $\gamma$. Computing the  ratio of the multistable region for each periodic window, we identify an increase of the ratio as $k$ also increases for weaker dissipation as, for greater dissipation, the ratio is almost constant. The basin stability and the basins of attraction show us that as $k$ increases, the area occupied by the periodic basin decreases and is restricted to larger values of the coordinate $p$ around the fixed points. These two results explain the incorrect measurement of shrimp-shaped properties, since the periodic behavior becomes more rare and more restricted to some initial conditions in the system.
    %, the scaling of the size of the periodic domains deviates from the obtained power law. 

\section*{Supplementary Material}

See the supplementary material for the data concerning the centers of the shrimp-shaped domains shown as red dots in Fig.~\ref{fig:paramspacetinyshrimp} and the coefficients of the polynomial fitting.

\section*{Declaration of competing interest}
    
The authors declare that they have no known competing financial interests or personal relationships that could have appeared to influence the work reported in this paper.

    \section*{Code availability}

    The source code to reproduce the results reported in this paper is freely available in the GitHub repository \cite{Rolim_Sales_dissipative-asymmetric-stdmap_2024}.

    \section*{Acknowledgments}

    This work was supported by the Araucária Foundation, the Coordination of Superior Level Staff Improvement (CAPES), the National Council for Scientiﬁc and Technological Development (CNPq), under Grant Nos. 309670/2023-3, 304398/2023-3, and by the São Paulo Research Foundation (FAPESP), under Grant Nos. 2018/03211-6, 2021/09519-5, 2022/12736-0, and 2023/08698-9. We would also like to thank the 105 Group Science (\url{https://www.105groupscience.com}) for fruitful discussions.

%\bibliography{refs}

%merlin.mbs aipnum4-1.bst 2010-07-25 4.21a (PWD, AO, DPC) hacked
%Control: key (0)
%Control: author (8) initials jnrlst
%Control: editor formatted (1) identically to author
%Control: production of article title (0) allowed
%Control: page (1) range
%Control: year (1) truncated
%Control: production of eprint (0) enabled
%

\end{document}